\documentclass[preprint,12pt]{revtex4}

\usepackage{amsthm}
\usepackage{graphicx}

\newcommand{\ket}[1]{\mbox{$ | #1 \rangle $}}

\begin{document}

\title{Simple, pulsed, polarisation entangled photon pair source}

\author{N. Bruno}
\author{E. Zambrini Cruzeiro}
\author{A. Martin}
\author{R. T. Thew}


\address{Group of Applied Physics, University of Geneva, Switzerland}


\begin{abstract}
We report the realisation of a fibred polarisation entangled photon-pair source at 1560 nm based on a type-II non linear interaction and working in the picosecond regime. By taking advantage of a set of filters, we deterministically separate the photons and project them into wavelength separable states. A standard entanglement measurement with a net interference visibility close to 1, proves the relevance of our approach as an enabling technology for entanglement-based quantum communication.
\end{abstract}



\maketitle

\section{Introduction}

Entanglement is a fundamental resource in quantum information science and sources of photonic entanglement are key enabling technologies for quantum communication~\cite{Gisin2007}. Entangled photons can be generated from a wide range of single-, or heralded-photon, sources~\cite{Eisaman2011}, or directly via interactions in materials with non linear optical susceptibility, such as spontaneous parametric down conversion (SPDC). A simple way to generate polarisation entangled photon pairs consists of employing a type-II non linear crystal~\cite{Kwiat1995}, which allows one to generate pairs of photons with orthogonal polarisation 
and then exploit the conservation of energy and momentum to produce entanglement. 

However, a constraint arises when we want to use a collinear configuration or waveguide crystals~\cite{Tanzilli2012}, for example, to increase the source brightness. The simplest approach is to use a balanced beam splitter, in which case the two photons take different paths half of the time. Hence, one immediately has only $50\%$ efficiency and the scheme only works in post-selection.  We can no longer deterministically separate the entangled photons to obtain polarisation entangled pairs, as they are indistinguishable in all the other degrees of freedom~\cite{Martin2010,Zhong2010}. A solution to this can be found by engineering two different  type II phase matching conditions, for non-degenerate pairs of photons, in the same crystal~\cite{Thyagarajan2009,Herrmann2011}. 

In this work, we report on an extension of a scheme proposed in Ref.~\cite{Kaiser2012}, which exploits energy conservation and degenerate photon pairs  in SPDC along with readily available fibre components such as dense wavelength division multiplexer (DWDM) filters. However, in our case, the picosecond pulsed regime allows one to generate narrowband photons and overcome synchronisation problems. We can also adapt the filtering bandwidth to obtain spectrally separable photons that are a requirement for more complex quantum communication tasks.

\section{Principle}

In the present work, the realisation of the entangled photon pair is, as previously mentioned, based on a type II spontaneous parametric down conversion (SPDC) process in a non linear crystal.
As proposed in Ref.~\cite{Kaiser2012}, in order to generate entangled photon pairs with minimised losses (\textit{i.e.} the collinear photons are deterministically separated), it is necessary to be in a degenerated configuration. A dense wavelength division multiplexer (DWDM) with a filter wavelength slightly detuned from the central one, has the role of separating the photons. At the two outputs, labelled $a$ (Alice) and $b$ (Bob), if all the distinguishabilities between the H and V polarised photons are erased, we obtain an entangled state of the form: 
\begin{equation}\label{eq:state}
\ket{\Psi}=\frac{1}{\sqrt{2}}\left[\ket{H_{+\delta \omega_f}}_a\ket{V_{-\delta \omega_f}}_b+\ket{V_{+\delta \omega_f}}_a\ket{H_{-\delta \omega_f}}_b\right]
\end{equation}
where $\delta\omega_f$ represents the frequency difference between the central filter frequency and the degeneracy frequency of the photon pairs. 

Parametric down conversion processes in non linear crystals are governed by energy and momentum conservation laws: 
\begin{equation}
\omega_p = \omega_i + \omega_s ; \qquad
\vec k_p = \vec k_i + \vec k_s + \frac{2 \pi}{\Lambda} \vec z
\end{equation}
where $\omega$ and $\vec k$ represent respectively the frequency and the wavenumber for the pump (p), signal (s), and idler (i)~\cite{Fujii2007}. $\Lambda$ is the crystal poling period employed to compensate the crystal dispersion. It was chosen to produce degenerate photon pairs at 1560\,nm. This non linear process is governed by the Hamiltonian:
\begin{equation}
H = c\int d\omega_s  \, d\omega_i \, \epsilon(\omega_s,\omega_i) \, \varphi(\omega_s,\omega_i) \, a^\dag(\omega_s) \, a^\dag(\omega_i) + \rm h.c.
\end{equation}
$\epsilon(\omega_s,\omega_i)$ and $\phi(\omega_s,\omega_i)$ are the pump pulse envelope and the phase matching function, respectively, which fix the energy conservation and the phase matching conditions. If the pump pulse is gaussian the first factor is given by $\epsilon(\omega_s,\omega_i) = exp\left[ -(\omega_i + \omega_s - \omega_p)^2/4 \Delta \omega_p^2\right] $, with $\Delta \omega_p$ the pump frequency bandwidth. The second factor, for a crystal of a length $L$, is given by: $\varphi(\omega_s,\omega_i) = {\rm sinc}(L(k_i + k_s + \frac{2 \pi}{\Lambda} - k_p))$.

\begin{figure}[h!]
\begin{minipage}{0.49\columnwidth}
\center
a)\\
\includegraphics[width=\columnwidth]{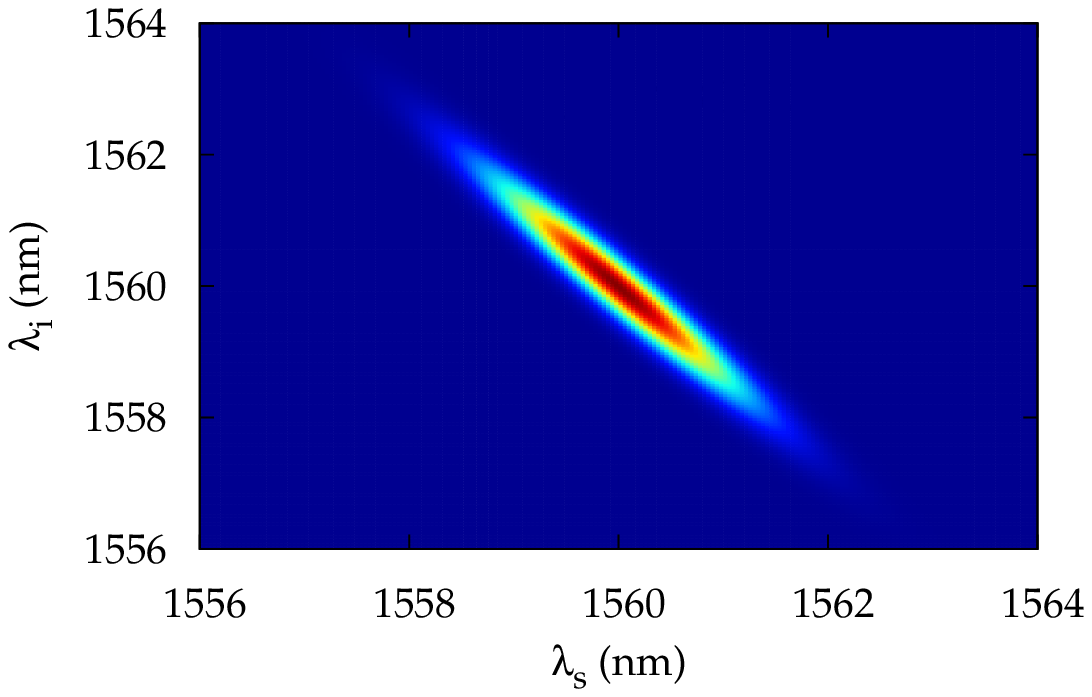}
\end{minipage}
\begin{minipage}{0.49\columnwidth}
\center
b)\\
\includegraphics[width=\columnwidth]{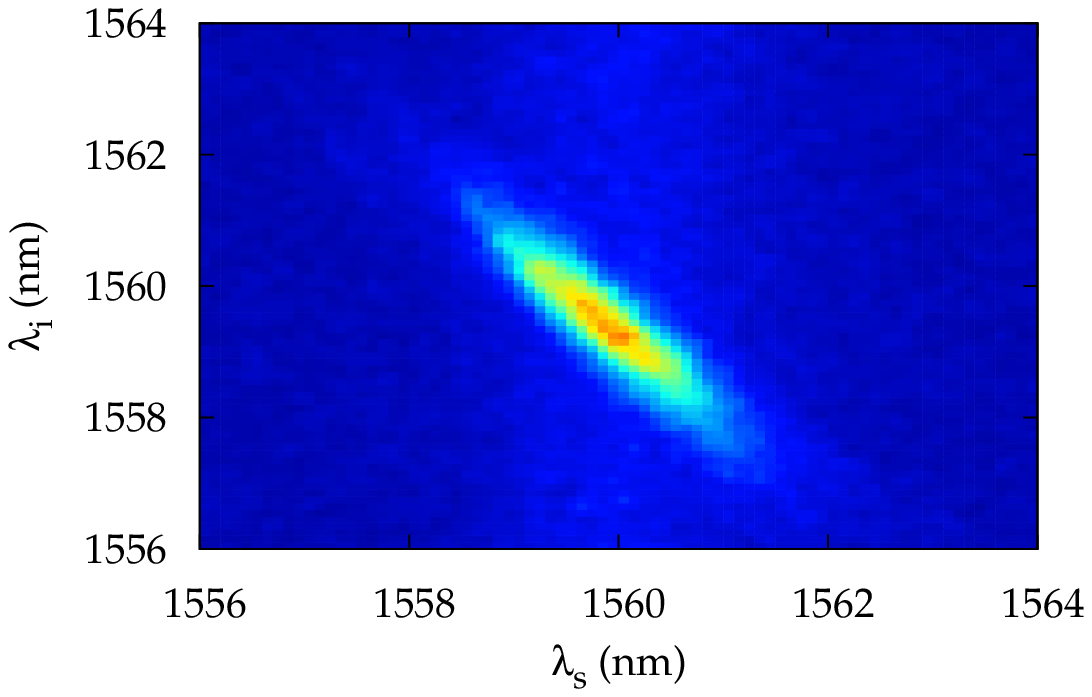}
\end{minipage}
\caption{\label{js}a)Numerical simulation of the joint spectral intensity for a 2 cm -long type II periodically poled Lithium Niobate (PPLN) crystal pumped by a 2 ps pulsed laser. b) Corresponding experimental measurement realised with two 200\,pm filters. Note that, signal and idler are associate to H and V polarisation, respectively.}
\end{figure}

 \figurename{~\ref{js}} represents the joint spectral intensity (JSI) ($J(\omega_s,\omega_i)=|\epsilon(\omega_s,\omega_i) \, \varphi(\omega_s,\omega_i)|^2$)  corresponding to our experimental configuration.  From \figurename{~\ref{js}} is possible to observe a correlation in wavelength, and, indeed in this case the state of the two photons at the output of the crystal is not separable in frequency. This state can be made spectrally separable by filtering one photon down to 200\,pm~\cite{Bruno2013a}.

\begin{figure}[h!]
\begin{minipage}{0.49\columnwidth}
\center
a)\\
\includegraphics[width=0.8\columnwidth]{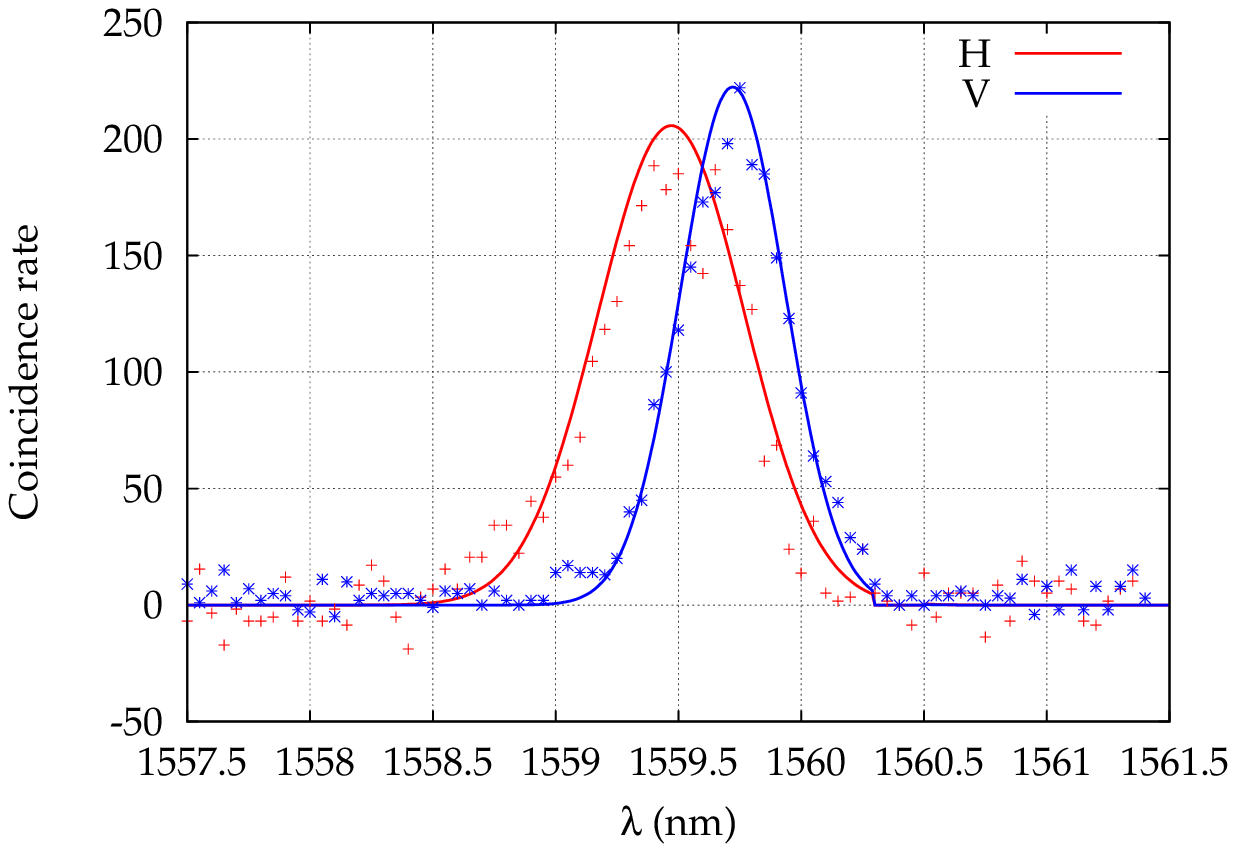}
\end{minipage}
\begin{minipage}{0.49\columnwidth}
\center
b)\\
\includegraphics[width=0.8\columnwidth]{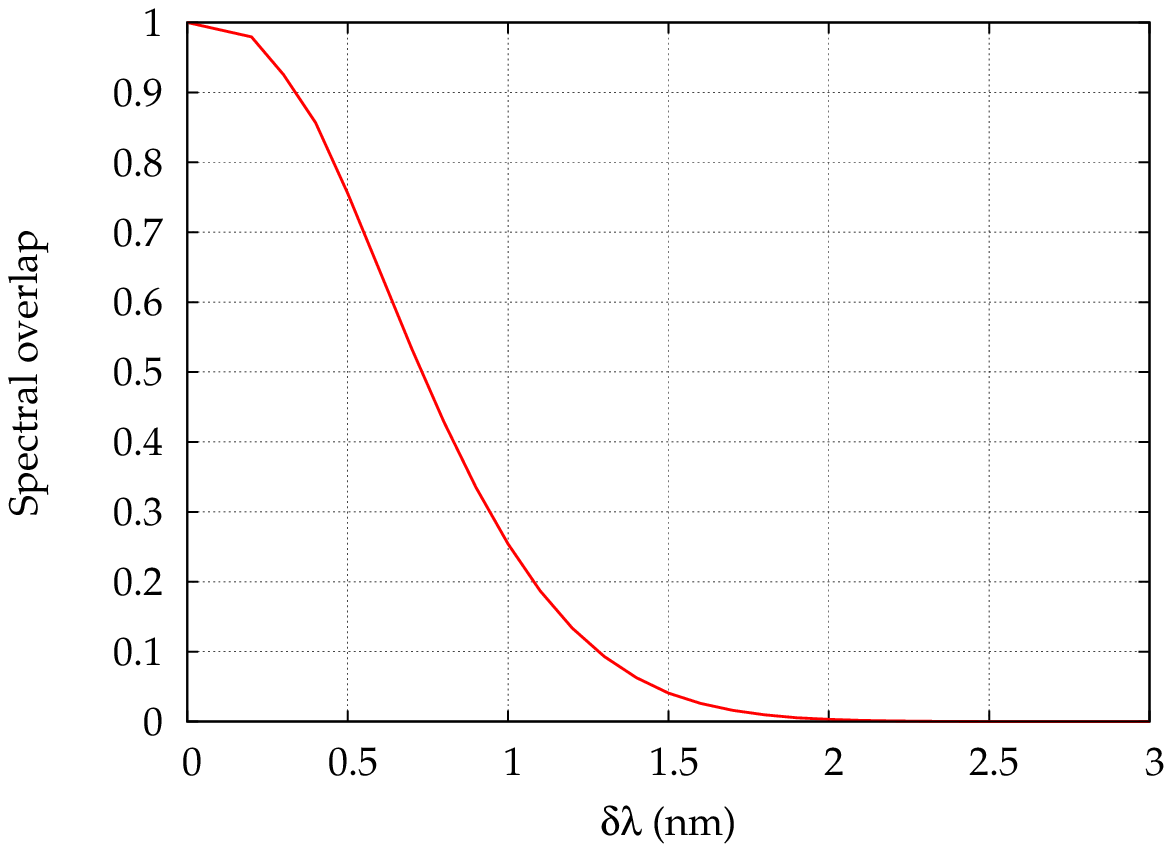}
\end{minipage}
\caption{\label{fig_overlap}a) Coincidence spectrum heralded by the photon passing the 0.2\,nm filter. The lines and the points represent respectively the theoretical prediction and the experimental results. b) Overlap in frequency between the two polarisation modes as a function of $\delta \lambda$ the difference between the degeneracy wavelength and the central wavelength of the 0.2\,nm filter.}
\end{figure}
 
If we include the action of such a filter in the JSI function, we can define the wavelength distribution of Bob's photons that are correlated with the photons sent to Alice. \figurename{~\ref{fig_overlap}a} gives an example of the wavelength distribution of Bob's photons. H and V polarised photons are distinguishable, which will clearly reduce the entanglement visibility~\cite{Grice1997}. The visibility in the diagonal basis is given directly by the overlap in frequency of the two polarisation modes. \figurename{~\ref{fig_overlap}b} shows this overlap as a function of the relative position of the filter compared to the degeneracy wavelength.

This distinguishability is not observed in the CW regime~\cite{Kaiser2012}, due to perfect spectral correlation between the photons. In the pulded regime the JSI contour is no longer an ellipse at 45$^\circ$ (see \figurename{\ref{js}}). The "tilt" angle $\theta$ of the phase matching function is given by:
\begin{equation}
\tan\theta=-\frac{k'_p-k'_{s}}{k'_{p}-k'_{i}}
\end{equation}
where $k'=\frac{d k}{d\omega}$ is the first derivative of the wavenumber, which in this case corresponds to a $\theta=59.96^\circ$.

To avoid this distinguishability it is necessary to add a second filter on Bob's arm, with a bandwidth adapted to just select the part of the spectrum where the two photons, H and V, overlap.

\section{Experimental realisation}

\begin{figure}[h!]
\center
\includegraphics[width=0.8\columnwidth]{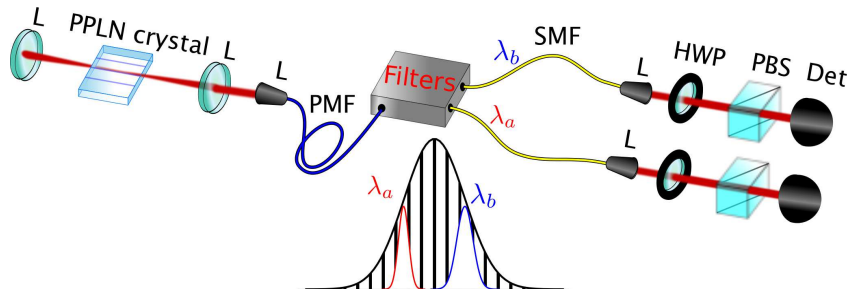}
\caption{Experimental setup. From left to right: pairs of orthogonally polarised photons degenerate at telecom wavelengths are generated via SPDC in a PPLN crystal, pumped by a 780 nm pulsed laser. The two photons are coupled into a PM fibre, then a filter splits $\lambda_a$ and $\lambda_b$ in order to generate the polarisation entangled pair. Half wave plates (HWP) and polarising beam splitters (PBS) followed by single photon avalanche photodiodes constitute the measurement apparatuses. }
\label{fig_setup}
\end{figure}

A scheme of the experimental setup is depicted in \figurename{~\ref{fig_setup}}: the two photons are generated via SPDC in a PPLN bulk crystal (Covesion), 2\,cm-long, pumped by a 2\,ps pulsed laser at 780\,nm with a repetition rate of 76 MHz (Coherent Mira 900). Type II quasi phase matching generates two orthogonally polarised photons.

A birefringent medium, e.g. a polarisation maintaining single mode fibre (PMF)is used to compensate the temporal walk-off introduced by the LN birefringence between the two orthogonally polarised photons. The photon pairs are directly coupled into the PMF with an efficiency of 50\%, the slow and fast axis as well as the length of the fibre (1,44\,m) are adapted to compensate the temporal walk-off introduced by the crystal.
To separate the two photons, a circulator is placed at the output of the PMF, a fibre Bragg grating (FBG) with a  200\,pm bandwidth is connected at the second port of the circulator, thus reflecting the filtered light towards the third port.  The central wavelength of the filter is detuned by $\sim 0.5$\,nm compared to the photons degeneracy wavelength. This detuning is such that one and only one photon per pair is selected by the filter. A second narrow filter (FWHM = 200 pm), used to eliminate residual distinguishabilities, can be added in Bob's arm.

\section{Result}

\begin{figure}[h!]
\begin{minipage}{0.49\columnwidth}
\center
a)\\
\includegraphics[width=\columnwidth]{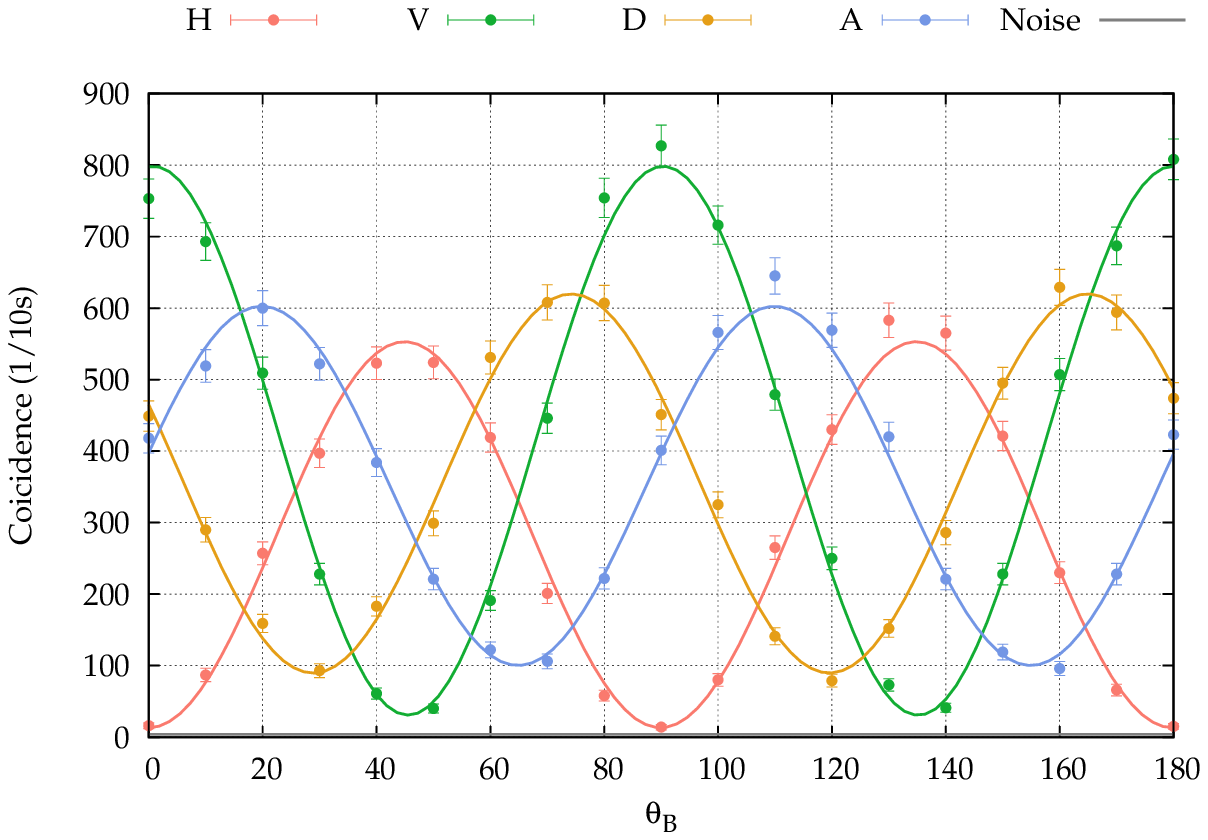}
\end{minipage}
\begin{minipage}{0.49\columnwidth}
\center
b)\\
\includegraphics[width=\columnwidth]{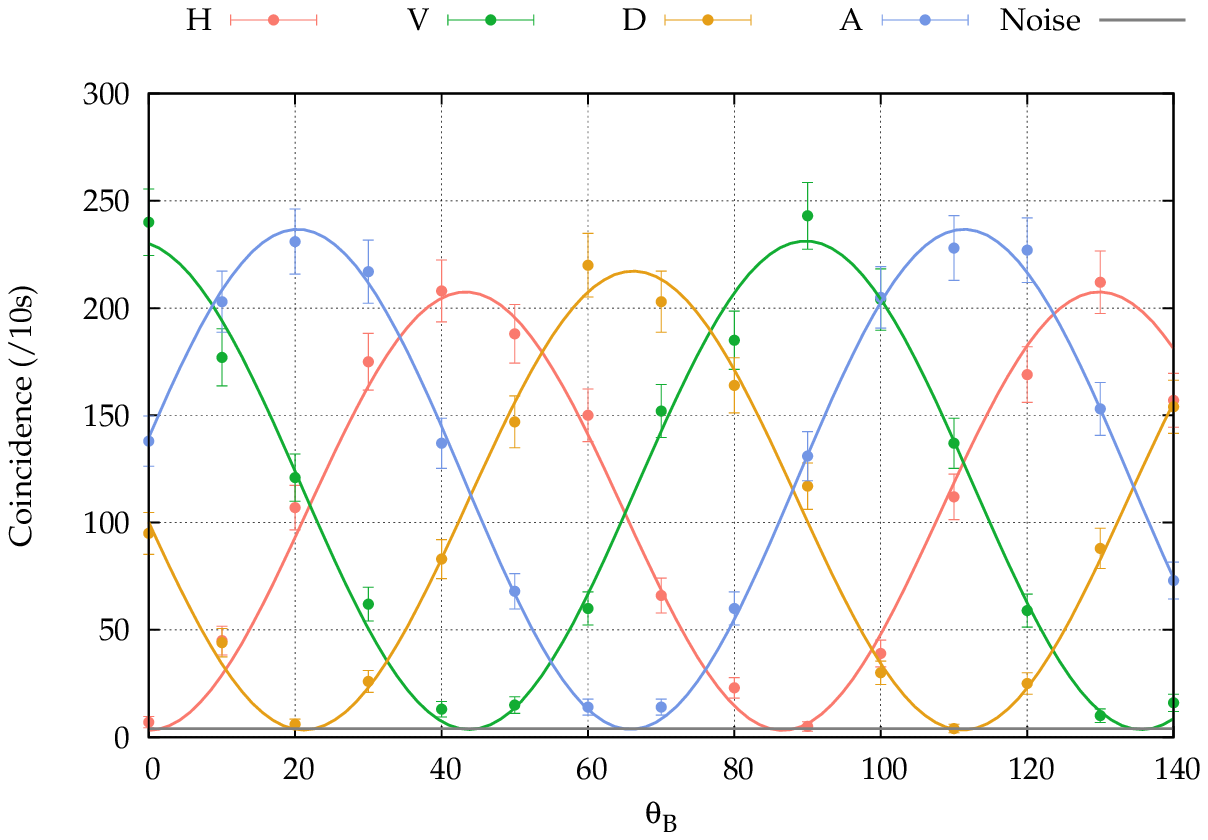}
\end{minipage}
\caption{\label{Bell_violation} Interference pattern for the standard entanglement measurement, \textit{i.e.} horizontal (H), vertical (V), diagonal (D), and anti-diagonal (A), for one filter and two filters configurations, (a) and (b), respectively. To reduce the probability of having double pairs, the pair creation probability was fixed at P = 0.01.}
\end{figure}

The photons reflected by the FBG are sent to Alice and the transmitted photons are sent to Bob, producing the state defined in equation~\ref{eq:state}. The polarisation analysers are composed by a half-wave plate, a polarising beam splitter and a gated InGaAs avalanche photodiode (APD, ID Quantique id210). A time-to-digital converter (TDC) records the coincidence counts between Alice and Bob for different polarisation bases.
To quantify the entanglement quality of the source, we perform a standard Bell inequality measurement. Alice's analyser is fixed to horizontal (H), vertical (V), diagonal (D) and anti-diagonal (A) positions, and the coincidences are recorded as a function of Bob's analyser angle ($\theta$). \figurename{~\ref{Bell_violation}a)} presents the experimental results. 
A high visibility is obtained in the \{H, V\} basis, as expected. However, in the \{D, A\} basis, the visibility is limited to 78\%, due to the fact that on Bob's channel there is some residual spectral distinguishability, as previously explained. This is related to the spectral properties of the photon pair, and it's a peculiar effect of the non linear material combined with the pump pulse shape and the filter shape.
Placing a second narrowband filter on Bob's arm allows one to improve the visibility up to 98\% for both the natural creation basis (HV) and the diagonal (DA) basis.

Table\ref{tab:vistable} reports the visibilities in all the four basis and the Bell parameter S for both configurations (without and with the filter on Bob' side). \figurename{~\ref{Bell_violation}.b)} shows the experimental data.

\begin{table}[!h]
\begin{center}
\begin{tabular}{c|c|c|c|c||c}
& H & V & D & A & S\\\hline
no filter & 98.9 (3.7)\% & 96.5 (2.7)\% & 78.0 (2.2)\% & 74.5 (3.1)\% & 2.46 (12)\\
filter & 98.9 (5.3)\% & 99.1 (4.8)\% & 98.0 (3.6)\% & 98.3 (3.3)\% & 2.79 (17)
\end{tabular}
\end{center}
\caption{Summary of the interference pattern visibility for the different orientations of the Alice's polarisation analyser and relative Bell parameter (S) for the configuration without and with the filter on Bob's arm.
\label{tab:vistable}}
\end{table}

\section{Conclusion}

Currently, the filtering is the major disadvantage with such an approach. The need for a second filter on Bob's arm doesn't arise in CW regime as the correlations in frequency between the two photons is perfect, i.e. the tilt angle of the JSI is 45. To overcome the double filtering in the pulsed regime, one could look at engineering the phase matching conditions in such a way that the JSI orientation is diagonal, which would lead to a perfect overlap between the entangled photons' spectra. In either case, the filtering can be easily replaced by ultra-DWDMs, which have a bandwidth of 25GHz, which is well suited to such a system.

We have demonstrated a simple way to generate polarisation entangled photon pairs based on collinear type-II PPLN phase matching in the pulsed regime. The high quality of the entanglement  is demonstrated by visibilities, in both bases, above $98\%$. The pulsed regime also allows one to have temporally well-defined photons for synchronisation with other sources and the filters allow one to deterministically separate the pairs and erase the frequency correlations. These characteristics, together with the simplicity and the intrinsic stability of the setup, make this approach a good candidate for future applications in entanglement-based  quantum communication networks.

\section*{Acknowledgement}

The authors thank B. Sanguinetti and Hugo Zbinden for help and fruitful discussions. Financial support from the Swiss NCCR - Quantum Science and Technology (QSIT).

\end{document}